\documentclass[trackchanges, twocolumn]{aastex701}

\begin{document}

\title{The Stack Search Tests on FAST Data: Discovery of Six Faint Isolated Millisecond Pulsars in NGC~6517 and NGC~7078 (M15)}

\author[0009-0007-6396-7891]{Yinfeng Dai}
\affiliation{Institute for Frontier in Astronomy and Astrophysics \& Faculty of Arts and Sciences, Beijing Normal University, Zhuhai 519087, China}
\affiliation{School of Physics and Astronomy, Beijing Normal University, Beijing 100875, China}
\email{yfeng.dai@foxmail.com}

\author[0000-0001-7049-6468]{Xing-Jiang Zhu}
\affiliation{Institute for Frontier in Astronomy and Astrophysics \& Faculty of Arts and Sciences, Beijing Normal University, Zhuhai 519087, China}
\email[show]{zhuxj@bnu.edu.cn}

\author[0000-0001-7771-2864]{Zhichen Pan}
\affiliation{National Astronomical Observatories, Chinese Academy of Sciences, Beijing 100010, China}
\affiliation{Guizhou Radio Astronomical Observatory, Guizhou University Guiyang 550025, China}
\affiliation{CAS Key Laboratory of FAST, National Astronomical Observatories, Chinese Academy of Sciences, Beijing 100101, China}
\affiliation{College of Astronomy and Space Sciences, University of Chinese Academy of Sciences, Beijing 100049, China}
\email[show]{panzc@bao.ac.cn}

\author[0000-0003-0597-0957]{Lei Qian}
\affiliation{National Astronomical Observatories, Chinese Academy of Sciences, Beijing 100010, China}
\affiliation{Guizhou Radio Astronomical Observatory, Guizhou University Guiyang 550025, China}
\affiliation{CAS Key Laboratory of FAST, National Astronomical Observatories, Chinese Academy of Sciences, Beijing 100101, China}
\affiliation{College of Astronomy and Space Sciences, University of Chinese Academy of Sciences, Beijing 100049, China}
\email[show]{lqian@bao.ac.cn}

\author[0000-0002-2394-9521]{Li-yun Zhang}
\affiliation{College of Physics, Guizhou University, Guiyang 550025, China}
\affiliation{International Centre of Supernovae, Yunnan Key Laboratory, Kunming 650216, China}
\email[show]{liy\_zhang@hotmail.com}

\author[0000-0001-6051-3420]{Dejiang Yin}
\affiliation{College of Physics, Guizhou University, Guiyang 550025, China}
\email{dj.yin@foxmail.com}

\author[0000-0001-7261-8297]{Yu Pan}
\affiliation{Chongqing University of Posts and Telecommunications, Chongqing, 40000, China}
\email{panyu@cqupt.edu.cn}

\author[0000-0003-0579-3014]{Bo Peng}
\affiliation{School of Information and Control Engineering, Southwest University of Science and Technology, Mianyang, 621010, China}
\email{pengball@swust.edu.cn}

\author[0009-0008-4109-744X]{Baoda Li}
\affiliation{College of Physics, Guizhou University, Guiyang 550025, China}
\email{gs.bdli21@gzu.edu.cn}

\author[0009-0001-6693-7555]{Yujie Lian} 
\affiliation{Institute for Frontiers in Astronomy and Astrophysics, Beijing Normal University, Beijing 102206, China}
\affiliation{Department of Astronomy, Beijing Normal University, Beijing 100875, China}
\email{ylian@mpifr-bonn.mpg.de}

\author[0009-0007-6770-5899]{Yaowei Li}
\affiliation{School of Physics and Electronics , Hunan Normal University, Changsha 410081, China}
\email{yw_li22@qq.com}

\author{Yuxiao Wu}
\affiliation{Chongqing University of Posts and Telecommunications, Chongqing, 40000, China}
\email{amerysain@gmail.com}

\author{Menglin Huang}
\affiliation{National Astronomical Observatories, Chinese Academy of Sciences, Beijing 100010, China}
\email{huangmenglin@nao.cas.cn}

\author{Qiaoli Hao}
\affiliation{National Astronomical Observatories, Chinese Academy of Sciences, Beijing 100010, China}
\email{qlhao@nao.cas.cn}

\author{Xingyi Wang}
\affiliation{National Astronomical Observatories, Chinese Academy of Sciences, Beijing 100010, China}
\email{wangxingyi@nao.cas.cn}

\author{Xianghua Niu}
\affiliation{National Astronomical Observatories, Chinese Academy of Sciences, Beijing 100010, China}
\email{xhniu@nao.cas.cn}

\author{Jinyou Song}
\affiliation{National Astronomical Observatories, Chinese Academy of Sciences, Beijing 100010, China}
\email{songjinyou@nao.cas.cn}

\author{Minglei Guo}
\affiliation{National Astronomical Observatories, Chinese Academy of Sciences, Beijing 100010, China}
\email{guominglei@nao.cas.cn}

\author{Shuangyuan Chen}
\affiliation{National Astronomical Observatories, Chinese Academy of Sciences, Beijing 100010, China}
\email{sychen@nao.cas.cn}

\begin{abstract}

We report the discovery of six faint millisecond pulsars (MSPs) in the globular clusters NGC~6517 and NGC~7078 (M15) using the Five-hundred-meter Aperture Spherical radio Telescope (FAST).
These discoveries were enabled by stacking power spectra from multiple observations, a method that effectively boosts the signal-to-noise ratio of faint sources.
In NGC~6517, we identified four new MSPs (NGC~6517S–V) with spin periods ranging from 3.68 to 6.02 ms and dispersion measures (DMs) between 182.45 and $182.85~{\rm pc\,cm^{-3}}$. 
In M15, two additional MSPs (M15M and M15N) were discovered, with spin periods of 4.83 and 9.28 ms, and DMs of 67.89 and $66.65~{\rm pc\,cm^{-3}}$, respectively.
A phase-coherent timing solution has been obtained for M15M; however, sparse detection rates currently preclude phase-connected solutions for the remaining five pulsars.
Current timing parameters suggest all six MSPs are isolated, which is consistent with the expected pulsar populations in core-collapsed globular clusters.
Notably, pulsars M15N, NGC~6517U, and NGC~6517V eluded detection by standard frequency-domain searches (e.g., PRESTO-based) and the Fast Folding Algorithm, demonstrating that the stack search technique significantly enhances detection sensitivity to inherently faint pulsar signals.

\end{abstract}

\section{Introduction} \label{sec:intro}

Globular clusters (GCs) are characterized by exceptionally high central stellar densities (typically exceeding $10^4$~M$_\odot$~pc$^{-3}$) and frequent dynamical interactions \citep[e.g.,][]{1996AJ....112.1487H}, making them highly efficient environments for the formation of millisecond pulsars (MSPs) via binary recycling and stellar exchange processes \citep[e.g.,][]{1982Natur.300..728A,2015aska.confE..47H}.
Since the discovery of the first GC pulsar in M28 \citep{1987Natur.328..399L}, the known population has grown significantly. 
As of March 2026, 359 pulsars have been discovered across 46 GCs, with approximately 93\% classified as MSPs and 57\% residing in binary systems\footnote{\url{https://www3.mpifr-bonn.mpg.de/staff/pfreire/GCpsr.html}}. 
These fractions are much higher than those observed in the Galactic field, 
reflecting the long evolutionary timescales and enhanced dynamical interactions intrinsic to GC environments \citep{1975AJ.....80..809H}.

While telescope sensitivity is a primary factor in pulsar discoveries, detecting faint MSPs remains notoriously difficult due to steeply declining luminosity functions \citep{2013MNRAS.431..874C}, interstellar scintillation at low dispersion measures (DMs), and acceleration-induced smearing in binary systems. 
Traditional searches that rely on dedispersion and Fourier transformation of a single observation are generally sufficient for detecting relatively bright sources \citep{2001PhDT.......123R}. 
Time-domain searches, such as those employing the Fast Folding Algorithm (FFA), can offer higher sensitivity than standard frequency-domain methods because the signal energy is coherently integrated \citep{2020MNRAS.497.4654M}. 
Nevertheless, single-observation searches frequently fail to uncover faint MSPs, regardless of whether Fast Fourier Transform (FFT) or FFA techniques are used. 
In the Fourier domain, weak signal power is often diluted across harmonics or obscured by red noise.
Conversely, in the time domain, folded profiles may only reach marginal significance and are easily discarded during automated candidate sifting, particularly when baseline variations or radio frequency interference (RFI) are present.
Consequently, even when a weak pulsar signal is present, the corresponding candidates may fail to pass the standard automated sifting criteria because their signal-to-noise ratios (SNR) do not reach the detection threshold.

The power spectrum stacking method (hereafter referred to as the stack search) mitigates these limitations by incoherently combining the power spectra from multiple observations. 
Assuming uniform integration times across observations, this technique boosts the detection sensitivity such that the SNR scales as ${\rm SNR} \propto \sqrt{N_{\rm obs}}$, with $N_{\rm obs}$ being the number of observations  \citep{1993PhDT.........2A}. 
The efficacy of this approach relies on the spectral stability of the target source. 
For isolated MSPs, whose spin periods change by only nanoseconds per year, signals from different observations remain precisely aligned in the Fourier domain, enabling robust power stacking and substantial SNR increases \citep{2005AAS...20718308S}. 
However, for binary systems, orbital motion induces significant Doppler shifts, causing the signal power to drift across frequency bins or become smeared in the power spectrum. This spectral leakage severely hinders effective power accumulation, resulting in significantly lower detection rates for binaries compared to their isolated counterparts.

The stack search has a proven track record of detecting faint isolated pulsars in GCs. 
To date, this technique has yielded at least 9 isolated GC MSP discoveries.
These include M15D and M15E, detected with the Arecibo Telescope\footnote{M15A/B: standard dedispersed FFT (harmonic-summed) search; M15C: acceleration/jerk search for a binary; M15F/G/H: coherent multiple-day search.}\citep{1993PhDT.........2A}, as well as 47 Tuc aa and ab, found using the Parkes Telescope \citep{2016MNRAS.459L..26P}. Additionally, the Green Bank Telescope facilitated the discovery of five pulsars in Terzan 5: Terzan 5ag and 5ah \citep{2005AAS...20718308S}, along with Terzan 5aj through 5al \citep{2018ApJ...855..125C}.

In this study, the stack search was applied to archival data obtained with the Five-hundred-meter Aperture Spherical radio Telescope (FAST) \citep{nan2011five, jiang2020fundamental}.
These observations are part of the FAST Globular Cluster Pulsar Survey (GC-FANS) \footnote{\url{https://fast.bao.ac.cn/cms/article/65/}}, which has thus far discovered over 60 pulsars across 16 of the 45 GCs visible in the FAST sky \citep[e.g.,][]{2020ApJ...892...43W,2021ApJ...915L..28P,2025ApJS..279...51L}.
Core-collapsed clusters are particularly promising targets for GC pulsar searches, as observations and recent dynamical modeling suggest that they host large populations of isolated MSPs \citep{2024ApJ...961...98Y,2024ApJ...977L..42K}.
M15 and NGC~6517 are the only two core-collapsed GCs accessible to FAST.
NGC~6517 is one of the densest clusters in the FAST sky, with a central luminosity density of $\rho_0=10^{5.29}~L_{\odot}~{\rm pc}^{-3}$. 
Previous FAST observations expanded the known pulsar population in NGC~6517 from 4 to 17 (comprising 16 isolated MSPs and 1 binary; \citealt{2021RAA....21..143P,2021ApJ...915L..28P,2024ApJ...969L...7Y}). 
Similarly, FAST previously uncovered four new pulsars in M15 (M15I--M15L; e.g., \citealt{2021ApJ...915L..28P,2024ApJ...974L..23W}), bringing its total known population to 12 (11 isolated and 1 binary).

In this work, six additional pulsars were discovered by reprocessing FAST archival data of NGC~6517 and M15 using the stack search.
The remainder of this paper is structured as follows: Section~2 describes the data and our reduction procedures. 
Section~3 presents the search results. 
Discussion is provided in Section~4, and Section~5 presents the conclusions.

\section{Data and Data Reduction} \label{sec:obs}
\subsection{Data}

This work is only based on FAST archival data.
A total of 23 archival FAST observations of NGC~6517, 
spanning from June 2019 to March 2024 (0.5–2.5~h per epoch), were used.
The 19 observational data of M15 were conducted between September 2019 and February 2024 (0.5–4.5~h per epoch).
The data were obtained only with the central beam of the FAST 19-beam receiver (covering 1.0 to 1.5 GHz),
with 4096 frequency channels (each channel is for 0.122 MHz) and a sampling time of 49.152 $\mu$s.
The total observation time is 36.43 and 44.04 hours for NGC~6517 and M15, respectively.
Most observations were performed in Tracking mode, 
except for the M15 observation on January 2, 2022, 
which was conducted in Snapshot mode.
Under this mode, the telescope will observe 4 positions to cover $\rm \sim$20$'$ region,
and only the first quarter of the data was used.

\subsection{Data Reduction} \label{sec:red}

RFI in the data were flagged using the routine \texttt{rfifind}
(with a time interval of 2.0\,s)
from the \textsc{PRESTO} package\footnote{\url{https://github.com/scottransom/presto}} \citep{2001PhDT.......123R, 2002AJ....124.1788R, 2003ApJ...589..911R}. 
The data were then dedispersed with a set of trial DM values with the \textsc{PRESTO} routine \texttt{prepsubband}.
The DM values of the previously detected pulsars were considered as minimum ranges.
For the NGC~6517, the DM range was set to 170--190~pc\,cm$^{-3}$ with a step of 0.05~pc\,cm$^{-3}$;
this range includes all previously known pulsars in the cluster (DM = 174.53--185.74~pc\,cm$^{-3}$; \citealt{2024ApJ...969L...7Y}).
For M15, the DM range of 63--70~pc\,cm$^{-3}$ with the DM step of 0.05~pc\,cm$^{-3}$, 
motivated by the DMs of previously known pulsars (DM = 65.59--67.73~pc\,cm$^{-3}$; \citealt{2024ApJ...974L..23W}).
During the dedispersion, the time series were referenced to the solar system barycenter.
Observations with longer integration times have higher frequency resolution, 
so these spectra from different observations cannot be stacked bin-by-bin.
Therefore, zeros were appended to the end of the dedispersed time series by
setting the parameter \texttt{-numout} in the routine \texttt{prepsubband}.
All dedispersed time series were subsequently zero-padded to a uniform duration of 2.5 hours for NGC~6517 and 4.5 hours for M15 observations, respectively.

The routine \texttt{realfft} was used to transform the time series to spectra with FFT \citep{1993PhDT.........2A}.
The routine \texttt{rednoise} from \textsc{PRESTO} was used to reduce the effect of the low-frequency interference signal (also known as red noise).
For each spectrum, 
the power spectrum and magnitude spectrum were computed from the real and imaginary parts as the squared modulus and the modulus, respectively.
The frequency range of these spectra was 0-10172~Hz, 
set by the Nyquist frequency for a sampling interval of 49.152~\textmu s \citep{oppenheim2009} in all of our GC pulsar survey observations.
Then, these spectra were stacked.


The detection sensitivity of pulsar signals can be improved by incoherently summing harmonics in frequency-based search technique \citep{2001PhDT.......123R}.
For pulsars with narrow pulse widths, harmonics even contain more power than the fundamental.
We therefore performed harmonic summing by folding the 2nd--16th harmonics onto the fundamental.

\subsection{Candidate Selection} 

These Candidates with statistical significance levels greater than or equal to $5 {\sigma}$ were selected from each spectrum following the harmonic summing ($\sigma$ denotes the standard deviation of the power values).
The stack search pipeline identified 65,456 and 3,385 candidates in NGC~6517 and M15, respectively.
Real pulsar signals are expected to be localized in DM and spin frequency: the detection statistic typically peaks near the best DM and declines away from it, 
while the recovered spin frequency remains constant across neighboring trial DMs \citep[e.g.,][]{2018ApJ...855..125C,2021RAA....21..143P}.
Thus, a DM-consistency filter was applied to filter out false positives, 
noting that RFI can persist over a broad range of trial DMs without a well-defined peak.
Given our trial-DM spacing of 0.05~pc\,cm$^{-3}$, we retained only candidates recovered in at least 10 trial-DM spectra (i.e., spanning $\geq$0.5~pc\,cm$^{-3}$), with detections occurring predominantly in adjacent DM trials.
The trial DM that maximizes the stacked power was considered as the optimal DM.
Following this filtering procedure, 2,638 and 2,443 candidates were obtained in NGC~6517 and M15, respectively.

Most RFI differs from pulsar signals in that the detection statistic shows no coherent dependence on DM and often appears scattered across trial DMs.
Thus, most RFI candidates were rejected through human-based inspection.
Some RFI can mimic pulsar-like DM behavior and survive this sifting.
So, for the remaining candidates, 
we extracted the trial frequencies and DMs and folded the dedispersed time series from all available observations using \texttt{prepfold} routine of \textsc{PRESTO}.
We visually inspected the folded profiles and diagnostic plots to identify potential pulsar signals.
Candidates showing pulse profiles in two or more observations were further investigated via timing analysis with \textsc{TEMPO}\footnote{\url{http://nanograv.github.io/tempo/}} \citep{2015ascl.soft09002N}.

As the result, all the previously known pulsars in NGC~6517 were detected and 4 new pulsars discovered in this cluster.
In M15, M15C was missed because it is currently very faint and has a very narrow pulse profile.
Two additional new pulsars were discovered in M15, namely M15M and M15N.

\begin{table*}[t]
\centering
\caption{Parameters of pulsars identified by the stack search in NGC~6517 and M15. 
\label{tab:Search_result}}
\small
\renewcommand{\arraystretch}{1.05}
\begin{tabular*}{\textwidth}{@{\extracolsep{\fill}}lccccc}
\hline\hline
\textbf{Pulsar}  & \textbf{Frequency} & \textbf{Period} & \textbf{DM} & \textbf{SNR} & \textbf{Harmonics} \\
\textbf{in NGC~6517}& \textbf{(Hz)} & \textbf{(ms)} & \textbf{(pc\,cm$^{-3}$)} &  & \textbf{(2-16)} \\
\hline
J1801$-$0857A & 139.36085 & 7.17562 & 182.65 & 201.9 & 2\\ 
J1801$-$0857B & 34.52990 & 28.96041 & 182.50 & 90.9 & 4\\ 
J1801$-$0857C & 267.47269 & 3.73870 & 182.40 & 205.8 & 3\\ 
J1801$-$0857D & 236.60056 & 4.22653 & 174.45 & 96.2 & 3\\ 
J1801$-$0857E & 131.55005 & 7.60167 & 183.30 & 115.0 & 2\\ 
J1801$-$0857F & 40.17353 & 24.89201 & 183.95 & 77.0 & 2\\ 
J1801$-$0857G & 19.38311 & 51.59130 & 185.15 & 35.4 & 4\\ 
J1801$-$0857H & 177.21939 & 5.64272 & 179.65 & 50.3 & 6\\ 
J1801$-$0857I & 307.29749 & 3.25418 & 177.90 & 19.8 & 2\\ 
J1801$-$0857K & 104.26998 & 9.59049 & 182.40 & 18.3 & 2\\ 
J1801$-$0857L & 165.09024 & 6.05729 & 185.75 & 23.0 & 2\\ 
J1801$-$0857M & 186.67740 & 5.35683 & 183.30 & 14.5 & 2\\ 
J1801$-$0857N & 200.21703 & 4.99458 & 182.65 & 26.2 & 2\\ 
J1801$-$0857O & 233.25719 & 4.28711 & 182.50 & 25.6 & 6\\ 
J1801$-$0857P & 180.63227 & 5.53611 & 182.95 & 19.6 & 2\\ 
J1801$-$0857Q & 137.77659 & 7.25813 & 182.50 & 6.1 & 4\\ 
J1801$-$0857R & 151.85462 & 6.58525 & 182.70 & 15.9 & 2\\ 
J1801$-$0857S$^{\dagger}$ & 264.94576(1) & 3.77435739(2) & 182.45 & 14.3 & 2\\ 
J1801$-$0857T$^{\dagger}$ & 271.682396(9) & 3.680768(1) & 182.50 & 12.1 & 2\\ 
J1801$-$0857U$^{\dagger}$ & 166.05330(2) & 6.02216257(9) & 182.85 & 8.6 & 2\\ 
J1801$-$0857V$^{\dagger}$ & 219.85022(1) & 4.54855118(3) & 182.55 & 8.7 & 2\\ 
\hline\hline
\textbf{Pulsar}  & \textbf{Frequency} & \textbf{Period} & \textbf{DM} & \textbf{SNR} & \textbf{Harmonics} \\
\textbf{in M15}& \textbf{(Hz)} & \textbf{(ms)} & \textbf{(pc\,cm$^{-3}$)} &  & \textbf{(2-16)} \\
\hline
B2127$+$11A & 9.03632 & 110.66446 & 67.20 & 813.1 & 5\\ 
B2127$+$11B & 17.81484 & 56.13298 & 67.80 & 359.3 & 2\\ 
B2127$+$11D & 208.21176 & 4.80280 & 67.25 & 176.0 & 2\\ 
B2127$+$11E & 214.98739 & 4.65143 & 66.60 & 285.6 & 4\\ 
B2127$+$11F & 248.32118 & 4.02704 & 65.60 & 72.1 & 5\\ 
B2127$+$11G & 26.55325 & 37.66017 & 66.95 & 22.3 & 3\\ 
B2127$+$11H & 148.29329 & 6.74339 & 67.05 & 24.9 & 2\\ 
J2129$+$1210I & 195.22874 & 5.12220 & 67.60 & 11.7 & 2\\ 
J2129$+$1210J & 84.44172 & 11.84249 & 66.55 & 23.8 & 3\\ 
J2129$+$1210K
& 1.03710 & 964.22736 & 66.70 & 21.1 & 15\\ 
J2129$+$1210L
& 1.00992 & 990.17819 & 67.50 & 17.8 & 8\\ 
J2129$+$1210M$^{\dagger}$  & 206.77469911787(9) & 4.836181623120(2) & 67.897(8) & 13.6 & 6\\  
J2129$+$1210N$^{\dagger}$  & 107.64786(1) & 9.2895482(1)
 & 66.65 & 10.2 & 4\\
J2129$+$1210O$^{\ddagger}$ & 90.35980 & 11.06687 & 67.50 & 11.0 & 5\\
\hline\hline
\end{tabular*}
\tablecomments{
The frequency and period of the newly discovered pulsars were fitted using \textsc{TEMPO}.
J1801$-$0857J is identified as a harmonic of J1801$-$0857B and is therefore omitted. 
B2127+11C, a short-orbit binary, was not detected in result and is therefore excluded from the table.
In addition, harmonic signals were detected from the long-period pulsars J2129$+$1210K ($P\approx 1928$ ms) and J2129$+$1210L ($P\approx 3960$ ms); the closest candidates correspond to their second and fourth harmonics, respectively.
Sources marked with $^{\dagger}$ denote the newly discovered MSPs reported in this work. 
Further details on J2129$+$1210O (marked with $^{\ddagger}$) are presented in \citet{2025RAA....25g1001D}.}
\end{table*}

\section{Results}

Based on the timing solutions from previous studies (\citealt{2024ApJ...969L...7Y} and \citealt{2024ApJ...974L..23W}), 
all previously known pulsars have been identified (Table~\ref{tab:Search_result}) except B2127+11C (M15C).   
The M15C is a double neutron star system with an orbital period of 8 hours \citep{1991ApJ...374L..41P}.
Notably, it was among the brightest pulsars in the cluster during the 1990s \citep{1990Natur.346...42A}.
Its signal disappeared in FAST data since 2022  \citep{2024ApJ...974L..23W}, 
and thus in the stacked spectrum the signal was averaged by many non-detection observations.
The spin frequencies and DM values of the known pulsars obtained from our detections 
are compared with the corresponding values in published timing solutions. 
The associated errors are of the order $10^{-4}$~Hz and  $0.2$~pc\,cm$^{-3}$, respectively.
Notably, J1801$-$0857Q (NGC~6517Q) has the lowest SNR (6.1) of all pulsars detected, due to its double-peaked profile with an inter-pulse; its second harmonic yields a much higher SNR of 16.0.

\begin{figure*}[!htb]  
\centering
\setlength{\tabcolsep}{20pt}

\begin{tabular}{cc}
  \textbf{(a) NGC~6517S} & \textbf{(b) NGC~6517T} \\
  \includegraphics[height=0.28\textheight]{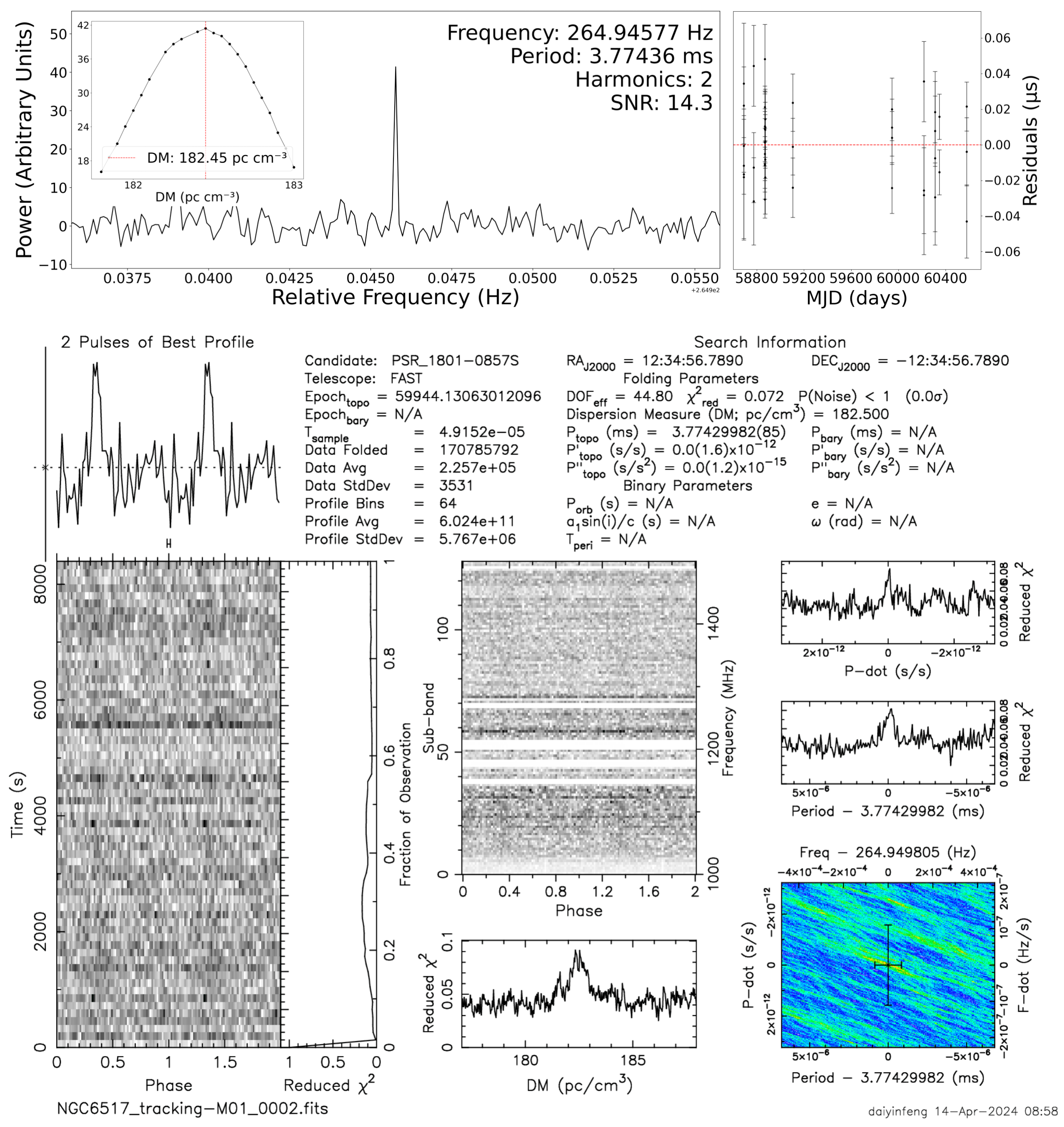} &
  \includegraphics[height=0.28\textheight]{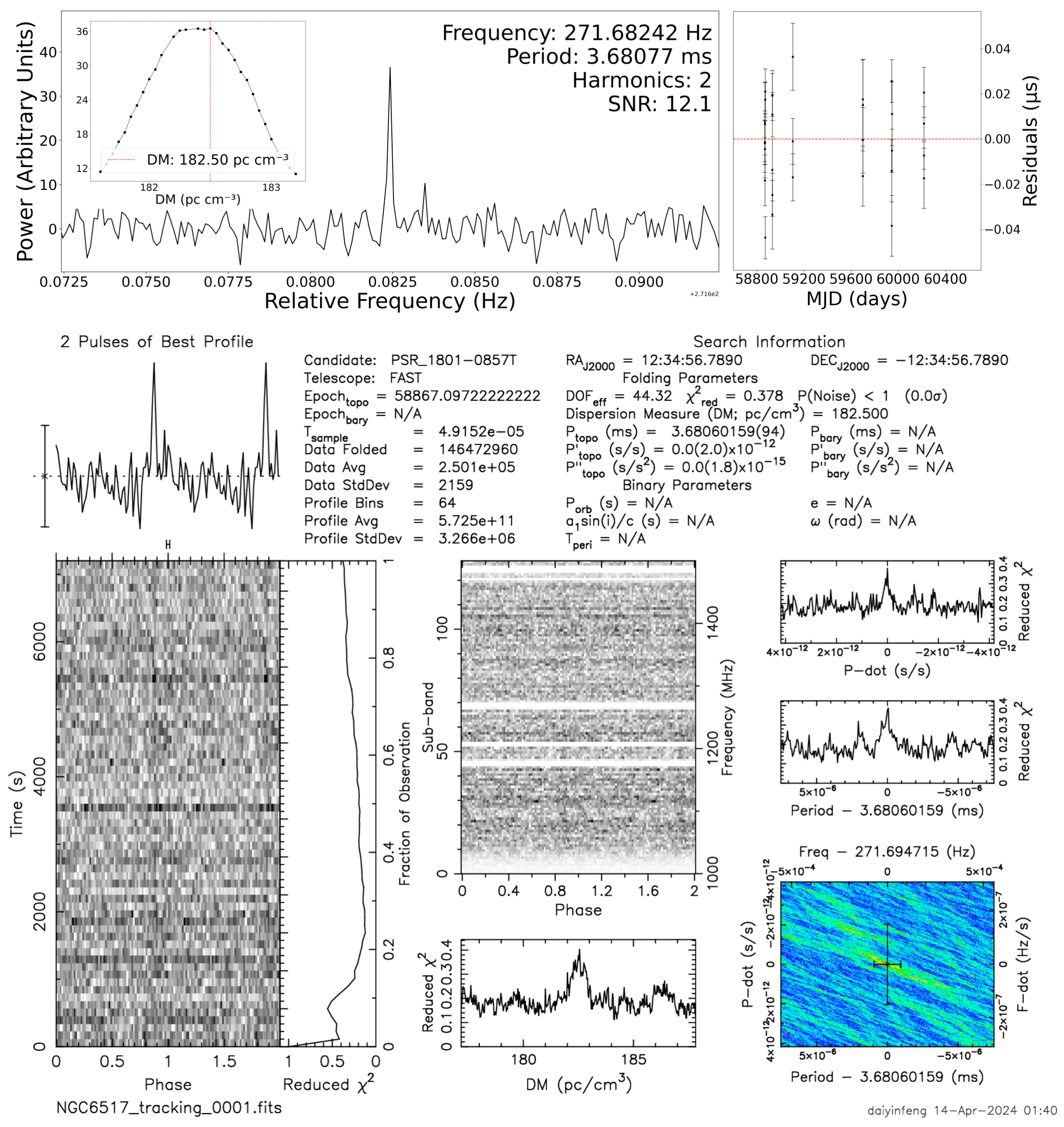} \\[0.2em]
  \textbf{(c) NGC~6517U} & \textbf{(d) NGC~6517V} \\
  \includegraphics[height=0.28\textheight]{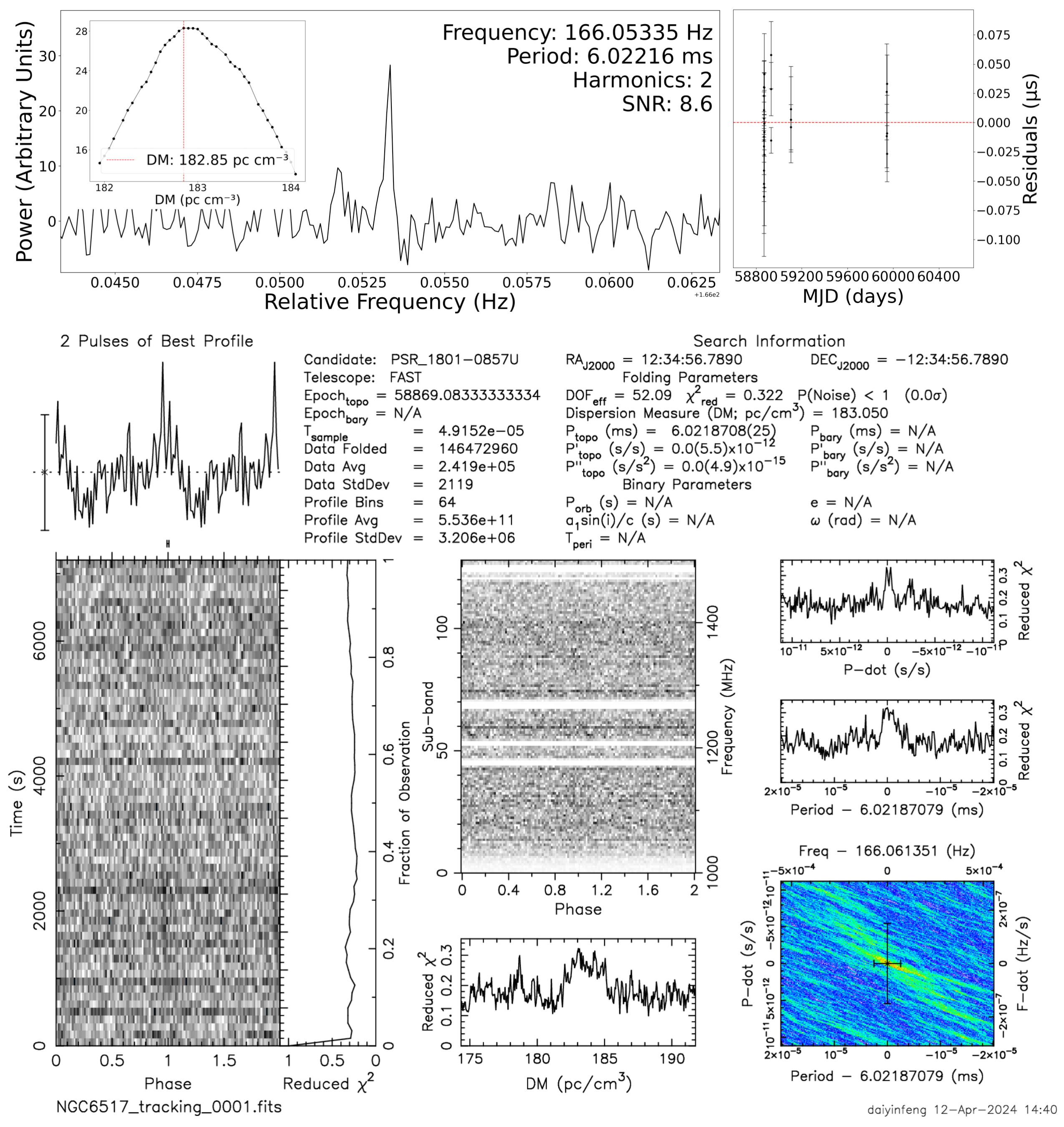} &
  \includegraphics[height=0.28\textheight]{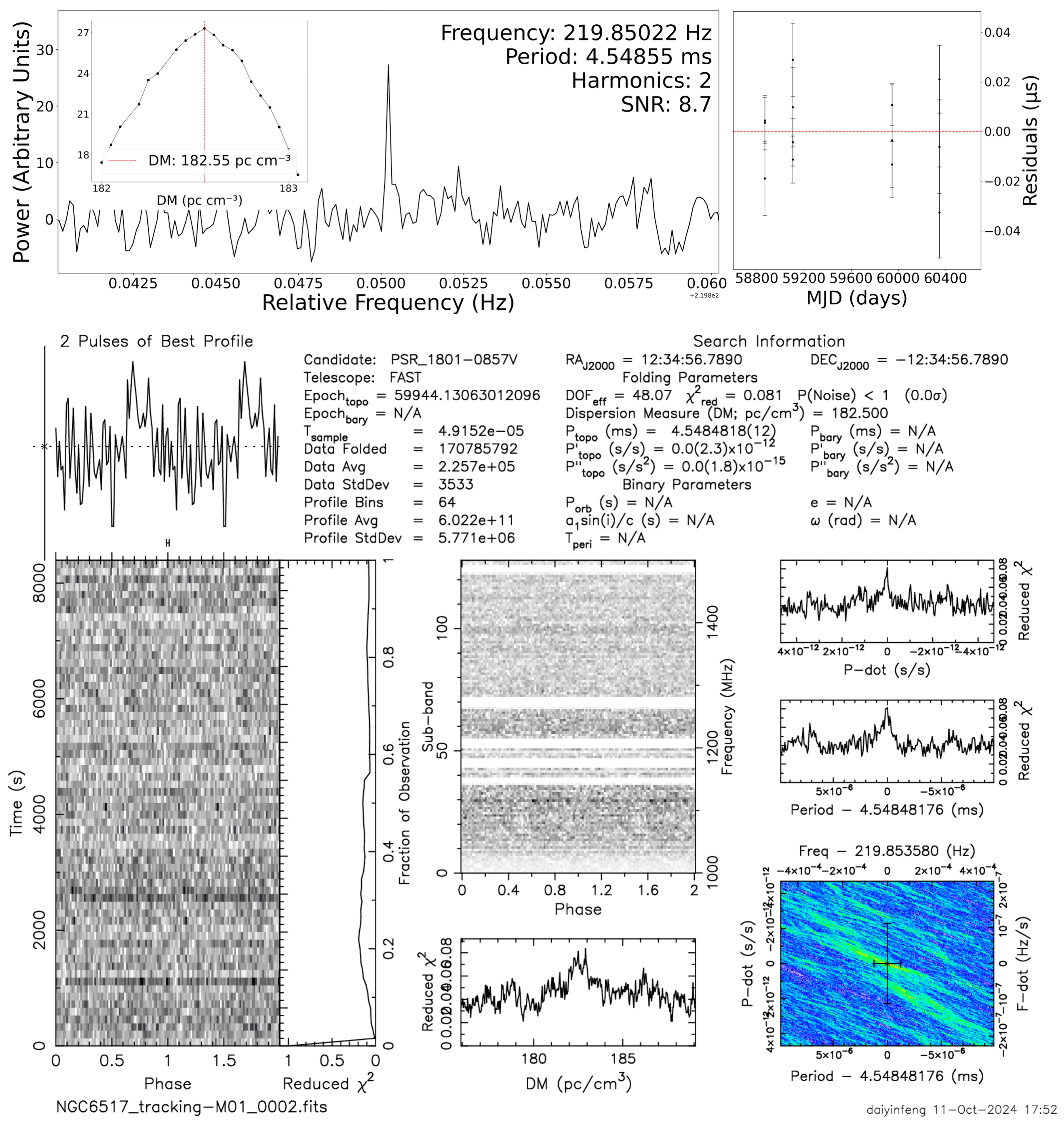} \\[0.2em]
  \textbf{(e) M15M} & \textbf{(f) M15N} \\
  \includegraphics[height=0.28\textheight]{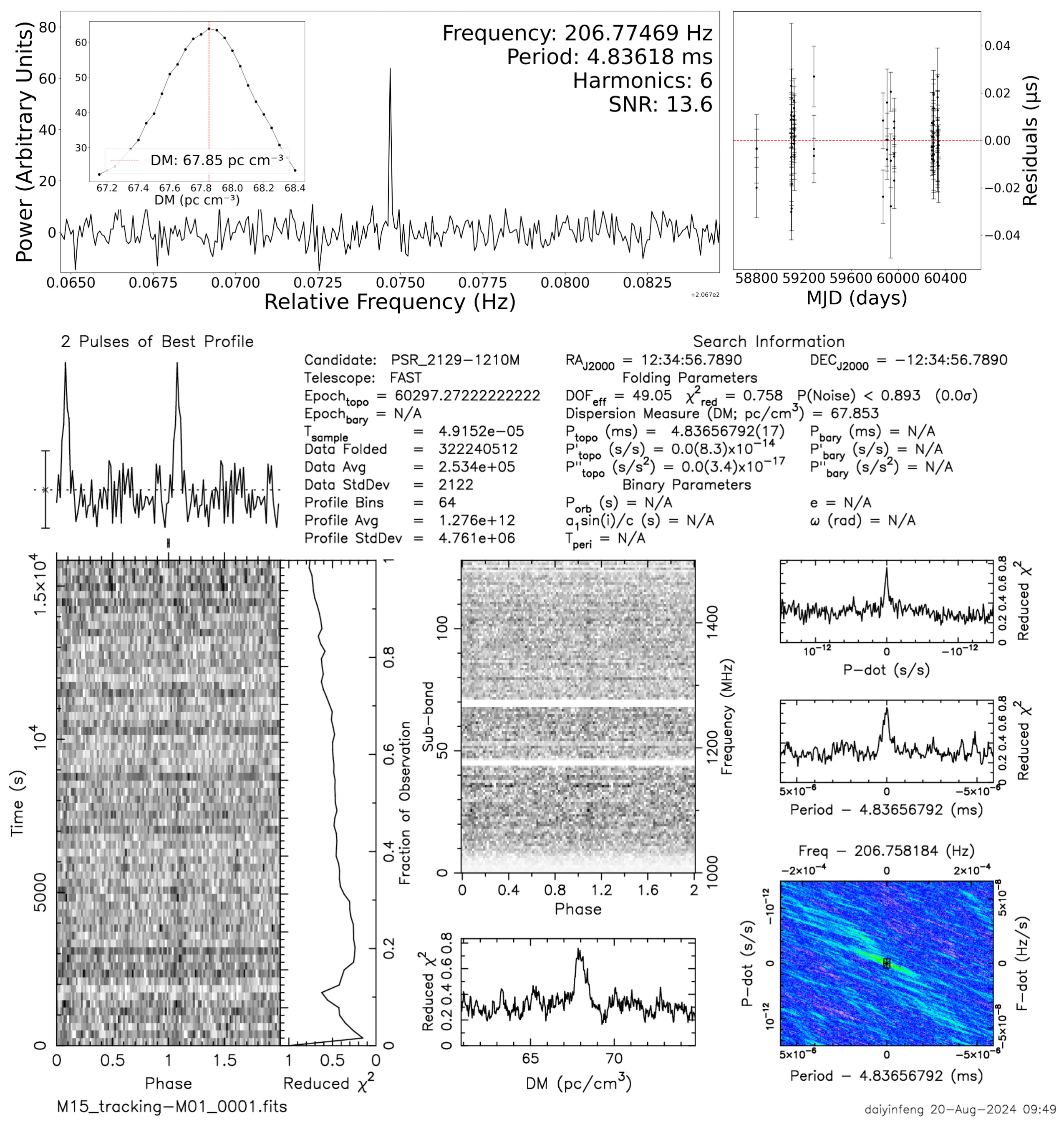} &
  \includegraphics[height=0.28\textheight]{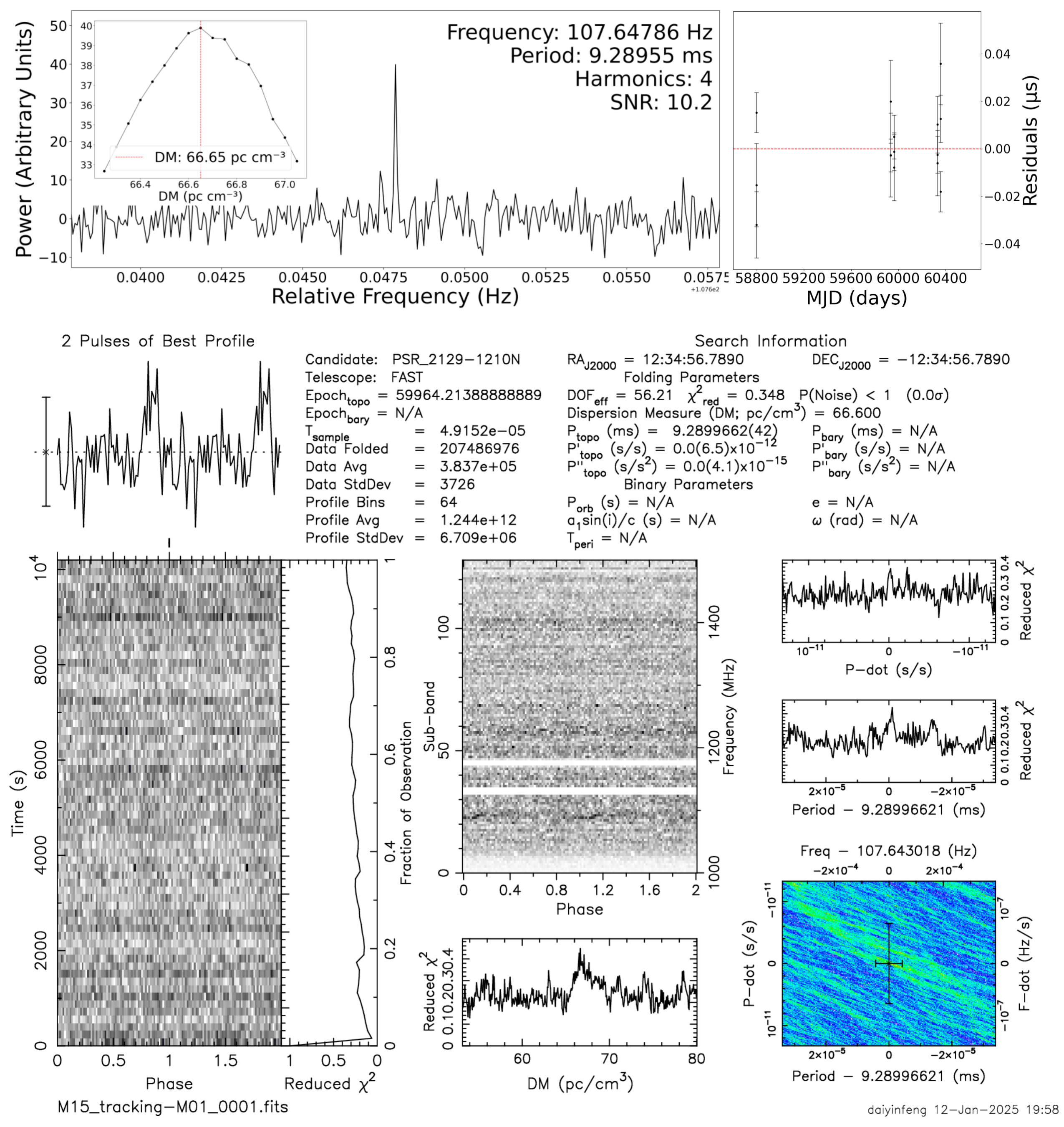} \\
\end{tabular}

\caption{Diagnostic plots for the six newly discovered pulsars in NGC~6517 and M15. Panels (a)--(d) show NGC~6517S/T/U/V, and panels (e)--(f) show M15M/N. In each panel, the upper subplots display the stacked power spectrum (left) and the timing analysis (with ``JUMPs'') residuals (right), and the lower subplots show the folded pulse profile and associated diagnostics.}
\label{fig:Convert}
\end{figure*}

Apart from the known pulsars, most candidates were identified as RFIs (e.g., appearing in only a single observation and/or lacking a clear broadband pulse signature). To confirm or reject the remaining candidates, a timing analysis was performed. Ultimately, only six optimal candidates were identified.
Among these six sources, only M15M has a phase-connected timing solution using the \textsc{TEMPO} package. 
Timing parameters derived for M15M are presented in Table~\ref{tab:M15M_Timing}.
For the other five pulsars, their intrinsic faintness means only spin frequencies were fitted (see Table~\ref{tab:Search_result}), with JUMPs between observations. They show highly consistent barycentric spin periods across multiple observations, thus providing strong timing constraints and indicating that they are likely isolated.
These discoveries are made possible by the high sensitivity of the stack search for isolated pulsars,
whereas if these are binaries, 
they are highly likely to be missed as orbital motion would broaden their signal peaks in the frequency spectrum.
The timing residuals versus MJD for all new pulsars are shown in the upper-right panel of Figure~\ref{fig:Convert}.


\subsection{New pulsars in NGC~6517}

Four new pulsars are discovered in NGC~6517,
namely J1801$-$0857S/T/U/V (hereafter NGC~6517S/T/U/V).
All of them have spin periods of $P<10$~ms.
Their detected DM values
are consistent with those of the previously known pulsars in NGC~6517 (mean DM $=182.1$~pc\,cm$^{-3}$).
Among the 23 observations used in our study, 
NGC~6517S/T/U/V were detected for 15, 12, 8, and 4 times, respectively,
based on the folding results (with SNR $>$ 7).
The discoveries increased the known pulsar population in NGC~6517 from 17 to 21; of these, 20 are classified as isolated MSPs \citep{2024ApJ...969L...7Y}, consistent with the high fraction of isolated pulsars expected in high-density, core-collapsed GCs \citep{2024ApJ...961...98Y}.

\subsection{New pulsars in M15}

Two new pulsars, namely J2129$+$1210M and J2129$+$1210N (hereafter M15M and M15N), were discovered in M15.
Both of them are isolated.
M15M was detected in 13 observations, 
enabling a phase-connected timing solution.
M15M has a spin period of $P=4.836$~ms and a DM value of 67.897(8)~pc\,cm$^{-3}$.
Its DM is the largest among all the M15 pulsars.
The timing position is $\alpha_{\rm J2000}=21^{\rm h}29^{\rm m}58\fs7175(4)$ and $\delta_{\rm J2000}=+12^{\circ}09'59\farcs69(1)$, located $6\farcs1\pm0\farcs9$ from the center of M15 in the sky plane \citep{2021MNRAS.505.5978V}.  

M15N was detected in 4 out of 19 observations.
Its low detection rate is insufficient for a phase-connected timing solution.
No significant orbital acceleration is detected in the available detections, suggesting that M15N is likely isolated.  
The optimal DM value of $66.65$~pc\,cm$^{-3}$ was determined by selecting the trial DM that maximized the detection significance across our full DM search range.
\begin{table}[t]
\centering
\caption{the timing parameters of new pulsar M15M.}\label{tab:M15M_Timing}
\small
\renewcommand{\arraystretch}{1.05}
\begin{tabular}{lc}
\hline\hline
Pulsar\dotfill  & J2129+1210M  \\
\hline
Right Ascension, $\alpha$ (J2000)\dotfill  & 21:29:58.7175(4)  \\
\hline
Declination, $\delta$ (J2000)\dotfill   & +12:09:59.69(1)  \\
\hline
Spin Frequency, $f$ (s$^{-1}$)\dotfill & 206.77469911787(9) \\
\hline
1st Spin Frequency derivative, $\dot{f}$ (s$^{-2}$)\dotfill & $1.0127(1)\times10^{-14}$   \\
\hline
Reference Epoch (MJD)\dotfill & 60297.272  \\
\hline
Start of Timing Data (MJD)\dotfill  & 58796.501  \\
\hline
End of Timing Data (MJD)\dotfill  & 60339.289  \\
\hline
Dispersion Measure, DM (pc cm$^{-3}$)\dotfill  & 67.897(8) \\
\hline
Number of TOAs\dotfill    & 89  \\
\hline
Residuals RMS ($\mu$s) \dotfill  & 50.47  \\
\hline
Detections\dotfill    & 15(19) \\
\hline\hline
\end{tabular}
\end{table}
\section{Discussion}

\subsection{Reasons of missing these new pulsars}

All six new pulsars presented in this work were missed by any previous pulsar search efforts.
Using the precisely measured spin periods and DMs for these sources, a retrospective analysis was performed to identify the exact stage of the standard search pipeline at which these signals were missed.
The M15M, 
which can be the brightest one among all the discoveries, 
was only detected once among 19 observations and with a given DM value.
Undoubtedly, this single detection will not pass the candidate sifting with the PRESTO script \texttt{ACCEL\_sift.py}.
Other pulsars were not detected for even once and thus missed.
As the FFA may have higher detection rates for these isolated MSPs \citep[e.g.,][]{2025ApJ...991...38L},
it was applied, too.
As the results, M15M and N, NGC~6517S, T, and U were detected by FFA, under the assumption that their spin periods were known in advance.
They were included in the candidate lists from FFA, however, 
it is worthy noting that they are quite faint so that they could easily be dismissed as noise by experienced researchers.
Maybe, the cross matching method (Yu et al. submitted) may have chance to catch them from the FFA candidate lists.
The NGC~6517V, which was identified with a lower SNR threshold, 
was only detected with the stack search.

\begin{figure}[t]
\centering 
\includegraphics[width=0.96\linewidth]{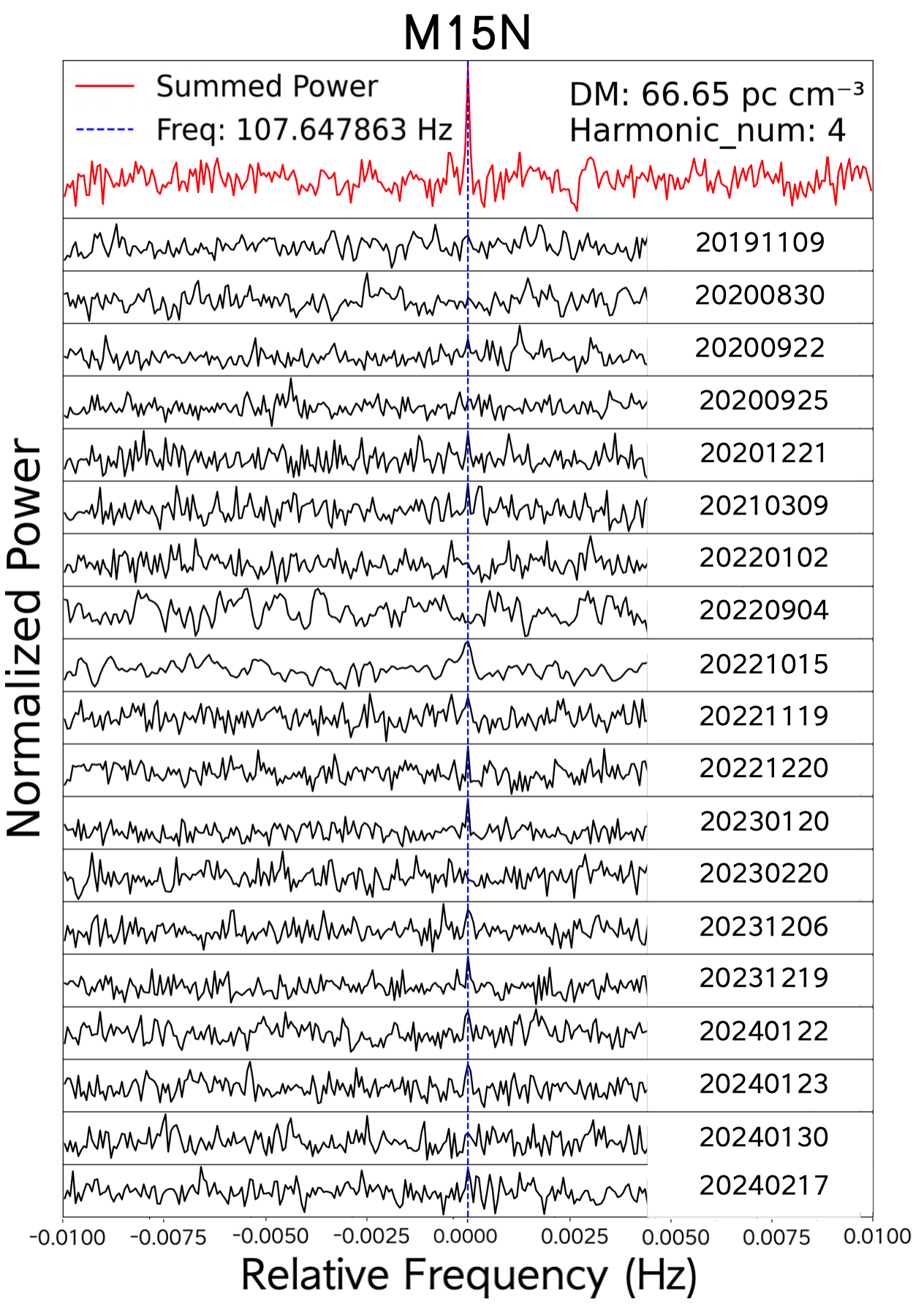}
\caption{Power spectra of M15N from 19 observations (panels with black solid lines, with observing dates) and the stacked spectrum (top panel with red solid line).
Every spectrum is with harmonic summing up to the 4th harmonic. 
The vertical dashed line marks the pulsar's spin frequency.}
\label{fig:M15N}
\end{figure}

On the other hand, the stack search offers two critical advantages: it not only boosts the SNR of faint pulsars, but also enables robust validation of signal stability through consistent detections across multiple observations, which is essential for distinguishing genuine astrophysical signals from RFI.
The spectrum of the signal across observations is plotted, using M15N as an example (see Figure ~\ref{fig:M15N}),
to test whether the detection is driven by one exceptional epoch or built from repeated sub-threshold appearances.
If a signal only appear once among many observations, this can be a highly nulling pulsar, or, more likely to be a RFI.
With the pulsar-like features, the signal of M15N is present at the same frequency in a substantial fraction of observations, 
demonstrating that the signal is persistent but typically below the single observation detection threshold.
As the spectra were stacked, the signal is obvious to be identified.

\subsection{Relationship between the number of observations and the SNR}
\label{subsec:snr_vs_nobs}

Assuming the pulsar signal does not scintillate (stable flux), the observations have the same integration time and settings, 
no RFIs or red noise affect the signal detection, and summing the same number of harmonics, 
when stacking the power spectra from observations, 
the signal power was added linearly ($\propto N_{\rm obs}$), 
whereas the noise increases as $\sqrt{N_{\rm obs}}$.
Theoretically, the detection significance in the stack search scales as ${\rm SNR}\propto \sqrt{N_{\rm obs}}$ \citep{1993PhDT.........2A,2020ARep...64..526T}.

Figure~\ref{fig:NGC6517_stack} illustrates how the signals of NGC~6517B, NGC~6517K, and NGC~6517V grow along the power spectra stacking. 
In the single spectrum ($N=1$; Figure~\ref{fig:NGC6517_stack}a) of each pulsar,
only the strongest source (NGC~6517B) shows a clear peak, 
while the weaker pulsars are marginal (NGC~6517K) or indistinguishable from the noise floor (NGC~6517V). 
After stacking 28 epochs ($N=28$; Figure~\ref{fig:NGC6517_stack}b), narrow peaks appear at the expected frequencies for all three pulsars, 
and their SNR increases substantially, 
demonstrating that the detections are built from repeated sub-threshold contributions rather than being dominated by a single exceptional observation.
An animation illustrating the stacking process is available as online supplementary material.

A total of 28 observations of NGC~6517 were used in this analysis.
For pulsar J1801$-$0857B (NGC~6517B), as it is in a binary system, the signal power from each observation could not accumulate at the same frequency bin.
Thus, its SNR increased by only a factor of $\sim$3.
Considering the variable integration times across the 28 observations and the intrinsic flux variability of MSPs \citep{lorimer2005handbook}, the $\sim$4-fold increase in SNR for NGC~6517K and V is deemed reasonable.

\subsection{Implications for Binary Pulsar Searches}

Currently the approach for finding binary pulsars involves requesting a much shorter or longer observing time than the orbit period.
The acceleration search \citep{2002AJ....124.1788R} introduces a non-zero spin frequency variation to describe a short segment of the orbital movement,
while the phase-modulation search needs $\rm \sim$1.5 or even more times \citep{2003ApJ...589..911R} of the orbital period as the observation time to obtain the orbital parameters.
While modern radio pulsar observations typically have integration times of several hours, the acceleration search method is widely employed and has identified numerous binary pulsars with orbital periods ranging from several hours to days.
On the other hand, there are very very few pulsars that are in 1-hr or even shorter orbit (e.g., M71E, 53 minute orbit, \citealt{2023Natur.620..961P}),
the phase-modulation search may not be suitable for a single radio observation.

\begin{figure*}[t]
\centering
\setlength{\tabcolsep}{8pt} 
\renewcommand{\arraystretch}{1.0}
\begin{tabular}{cc}
  \textbf{(a) Stack Before} & \textbf{(b) Stack After} \\ %
  \includegraphics[width=0.45\textwidth,keepaspectratio]{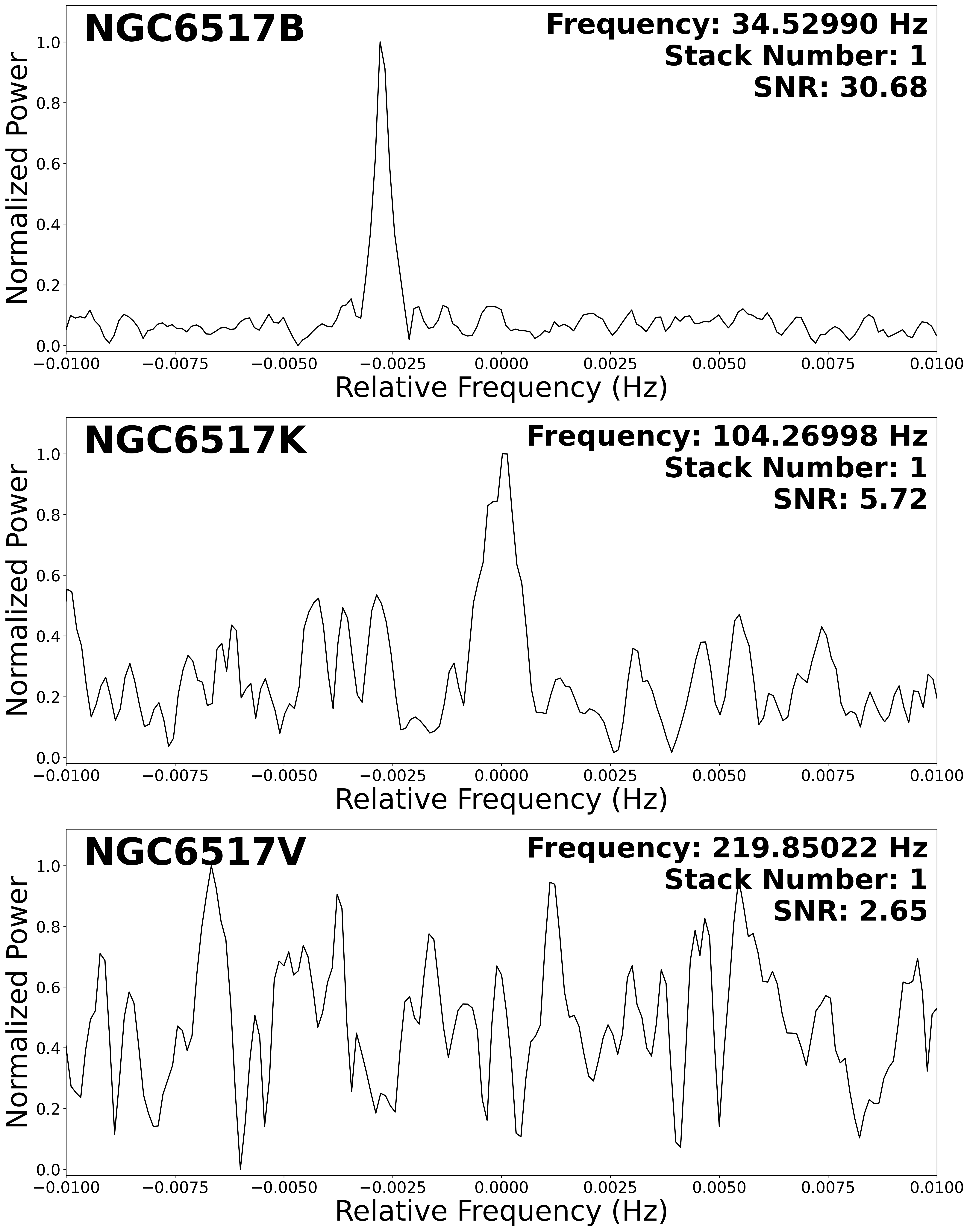} &
  \includegraphics[width=0.45\textwidth,keepaspectratio]{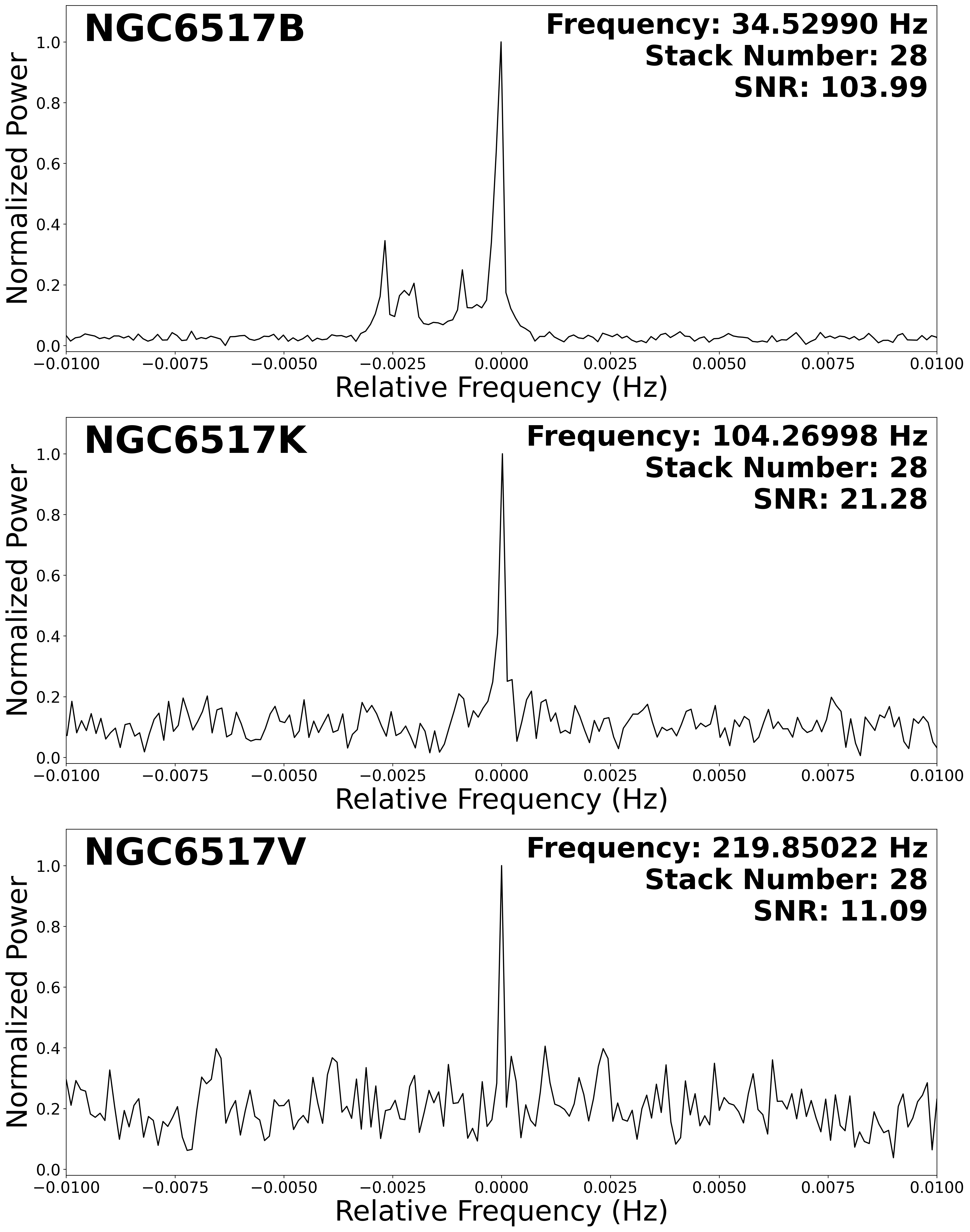} \\
\end{tabular}
\caption{Spectral stacking of the pulsar signal from 28 FAST observations of NGC~6517, including five additional recent 2.5-hr observations (with animation). 
(a) Power spectrum from a single 0.5-hr FAST observation.  
(b) Combined spectrum after stacking all 28 observations.  
The accompanying animation shows the stacking process for the 28 observations in the order of observation dates.
Note that NGC~6517B is a binary with an orbital period of 57.89 days, 
so its powers will not all sum at the same frequency.}

\label{fig:NGC6517_stack}
\end{figure*}

The stacked spectrum around the spin frequency of the known binary pulsar NGC~6517B ($P_{\rm b}\approx58.9$~day) exhibits a clear U-shaped (``horned'') envelope (Figure~\ref{fig:NGC6517_stack}b), 
which is qualitatively reminiscent of the sideband structure described by phase-modulation searches \citep{2003ApJ...589..911R}.
Thus, NGC~6517B may have chance to be detected by the phase-modulation search from the stacked spectrum.
The phase modulation search targets pulsars with very short orbital periods in long-integration observations (requiring $T_{\rm obs} \gtrsim 1.5 P_{\rm orb}$); 
when combined with the stack search, this algorithm can potentially detect binary pulsars with orbital periods spanning minutes to hours, and even days (e.g., NGC~6517B, which has a 58.9~day orbital period).
A dedicated implementation and quantitative assessment are beyond the scope of this Letter and will be explored in future work.

In addition, GCs may host multiple pulsars, making the stack search well-suited for analyzing GC observational data.
Similarly, this method is also efficient for specific classes of non-GC sources (e.g., high-energy point sources or supernova remnants).
The efficacy of the stack search for GCs is further enhanced by the narrow DM range of GC pulsars;
conversely, data processing will cost much time when the DM of potential pulsars in a GC is 
unknown—such as for M72 and M107, the two southernmost GCs in FAST sky (Peng et al., in prep).

\section{Conclusions}

An incoherent power spectrum stacking search was applied to archival FAST observations of NGC~6517 and M15, and discovered 6 MSPs: J1801$-$0857S-V (NGC~6517S-V) and J2129$+$1210M-N (M15M-N).
These MSPs exhibit spin periods ranging from 3.68 to 9.29 ms, with DM spanning 182.45 to 182.85~pc\,cm$^{-3}$ in NGC~6517, and 66.65–67.89~pc\,cm$^{-3}$ in M15.

Timing analyses indicate that all six of these pulsars are isolated systems.  Their signals are too weak to be robustly recovered in standard single-epoch searches, illustrating the advantage of stacking technique for detecting extremely faint isolated MSPs.
These discoveries increase the known pulsar populations in NGC~6517 and M15 by approximately 27\% and 18\%, respectively.
This suggests that a non-negligible fraction of the cluster pulsar population resides at very low luminosities, rendering them easily missed by traditional analyses based on individual observations.

The stack search improves detectability by accumulating signal power across multiple epochs while simultaneously averaging down stochastic noise, 
yielding a sensitivity gain proportional to $\sqrt{N_{\rm obs}}$.
By utilizing this approach, we successfully recovered all previously known isolated pulsars in both clusters and extended the capability of FAST searches for isolated MSPs.

Because several of these newly discovered pulsars are extremely faint, securing full timing parameters remains challenging. 
With the current dataset, a phase-connected timing solution has been achieved only for M15M. 
For the remaining five pulsars, the available detections are too sparse and the SNRs are too low to enable phase connection across different epochs.
Obtaining phase-connected timing solutions for these pulsars will require additional observations. 
Furthermore, the COBRA method \citep{2018MNRAS.473.5026L} can be adapted for single-fold input data and coherently combines all available observations (including non-detections) to derive precise timing solutions.
We will explore this technique in a future paper.
Overall, it is concluded that the stack search is a highly effective method for finding faint isolated MSPs, and it potentially holds promise for the future detection of compact binary pulsars.

\label{Conclusions}

\begin{acknowledgments}

This work is supported by the National Key R \& D Program of China No. 2022YFC2205202, No. 2020SKA0120100, NO. 2025SKA0140101 and the National Natural Science Foundation of China (NSFC, Grant Nos. 12373032, 12003047, 11773041, U2031119, 12173052, and 12173053). 
This work is also supported by Guizhou provincial natural science foundation project NO. ZD[2026]058.
Z.P. is supported by the CAS “Light of West China” Program and the Youth Innovation Promotion Association of the Chinese Academy of Sciences (ID 2023064).
L.Q. is supported by the Youth Innovation Promotion Association of CAS (ID 2018075, Y2022027) and the CAS “Light of West China” Program.  
X.-J. Zhu is supported by the National Key Research and Development Program of China (No. 2023YFC2206704), the Fundamental Research Funds for the Central Universities, and the Supplemental Funds for Major Scientific Research Projects of Beijing Normal University (Zhuhai) under Project ZHPT2025001.
We are grateful to Professor John K. Webb for his valuable advice and helpful comments during the preparation of this manuscript.
This work made use of the data from FAST (Five-hundred-meter Aperture Spherical radio Telescope) (https://cstr.cn/31116.02.FAST). 
FAST is a Chinese national mega-science facility, operated by the National Astronomical Observatories, Chinese Academy of Sciences. 
\end{acknowledgments}

\facilities{FAST}

\software{Tempo \citep{2015ascl.soft09002N}, PRESTO \citep{2001PhDT.......123R}}

\bibliography{Reference}
\bibliographystyle{aasjournalv7}

\end{document}